\PassOptionsToPackage{unicode}{hyperref}
\PassOptionsToPackage{hyphens}{url}
\documentclass[
  11pt,
]{article}
\usepackage{xcolor}
\usepackage[margin=1in]{geometry}
\usepackage{amsmath,amssymb}
\usepackage{iftex}
\ifPDFTeX
  \usepackage[T1]{fontenc}
  \usepackage[utf8]{inputenc}
  \usepackage{textcomp} 
\else 
  \usepackage{unicode-math} 
  \defaultfontfeatures{Scale=MatchLowercase}
  \defaultfontfeatures[\rmfamily]{Ligatures=TeX,Scale=1}
\fi
\usepackage{lmodern}
\ifPDFTeX\else
\fi
\IfFileExists{upquote.sty}{\usepackage{upquote}}{}
\IfFileExists{microtype.sty}{
  \usepackage[]{microtype}
  \UseMicrotypeSet[protrusion]{basicmath} 
}{}
\makeatletter
\@ifundefined{KOMAClassName}{
  \IfFileExists{parskip.sty}{%
    \usepackage{parskip}
  }{
    \setlength{\parindent}{0pt}
    \setlength{\parskip}{6pt plus 2pt minus 1pt}}
}{
  \KOMAoptions{parskip=half}}
\makeatother
\providecommand{\tightlist}{%
  \setlength{\itemsep}{0pt}\setlength{\parskip}{0pt}}
\usepackage[]{natbib}
\usepackage{graphicx}
\usepackage{bookmark}
\IfFileExists{xurl.sty}{\usepackage{xurl}}{} 
\hypersetup{
  pdftitle={The Khipu Problem: Institutional Legibility Under Distributed Cognition},
  pdfauthor={Krti Tallam KamiwazaAI krti@kamiwaza.ai},
  hidelinks,
  pdfcreator={LaTeX via pandoc}}

\title{The Khipu Problem: Institutional Legibility Under Distributed
Cognition}
\author{Krti Tallam\\
KamiwazaAI\\
krti@kamiwaza.ai}
\date{May 6, 2026}

\begin{document}
\maketitle

\section*{Abstract}\label{abstract}
\addcontentsline{toc}{section}{Abstract}

AI governance still tends to assume that the relevant object is a
bounded model or a bounded agent. That assumption is getting weaker.
Real systems increasingly distribute cognition across models, tools,
humans, context stores, retrieval layers, runtime policies,
authorization boundaries, and delegated institutional roles. In such
systems, the central governance problem is no longer only what the
system did, but whether later institutions can still read what the
system was
\citep{hollan2000distributed, selbst2019fairness, yao2022react, wu2023autogen}.

This paper introduces the \emph{khipu problem} for distributed AI: the
record can survive while the reading practice needed to interpret it
decays. Logs, traces, model versions, tool calls, outputs, and approval
artifacts may remain available while the institutional capacity to read
them as parts of one coherent cognitive episode disappears. We argue
that this failure is better understood as loss of \emph{interpretive
continuity} than as ordinary lack of observability.

The result is a distinct governance failure. Institutions must classify,
trust, audit, and constrain systems whose relevant identity is
distributed across components and whose legibility depends on
surrounding interpretive scaffolding. The problem is not merely missing
data. It is a structural mismatch between what can be represented and
what must still be decided under consequential conditions
\citep{kroll2017accountable, ananny2018seeing, tabassi2023airmf}.

We therefore argue that governance for distributed AI requires
preservation of interpretive continuity, not only trace retention. The
paper distinguishes missing evidence, ambiguous evidence, and
structurally unreadable evidence; argues that many consequential
outcomes are better understood as distributed cognitive episodes than as
bounded model outputs; and proposes governance workspaces together with
receipt-bearing governance surfaces as interpretive infrastructure for
preserving action identity, authority, boundary truth, evidential scope,
and consequential outcomes.

\section{Introduction}\label{introduction}

AI governance still tends to imagine a bounded object. The familiar
questions ask what a model did, what a model knew, whether a model was
safe, and what evidence about that model should be preserved for later
review. That frame remains visible even in work on documentation,
auditing, interpretability, and risk management
\citep{mitchell2019model, gebru2018datasheets, bender2018data, raji2020closing, doshivelez2017rigorous, tabassi2023airmf}.
But the technical substrate is moving. In consequential settings, the
relevant behavior is increasingly produced by coordinated episodes
across models, retrieval layers, tools, runtime policy, authorization
boundaries, and human intervention
\citep{yao2022react, schick2023toolformer, wu2023autogen, park2023generative, wang2023voyager, hong2024metagpt}.
The visible output may still appear as one answer or action. The
effective cognition that produced it is distributed.

That shift does not merely complicate attribution. It changes the object
that governance is trying to govern. A system can preserve immense
quantities of telemetry and still fail in a deeper way: it can retain
prompts, weights, tool calls, approvals, and outputs while losing the
institutional ability to read those artifacts as parts of one coherent
episode. What disappears is not the record, but the reading practice
that once made the record usable for judgment. This paper calls that
failure mode the \textbf{khipu problem}.

The name is deliberate. The analogy to the Inca khipu is meant to
isolate a specific governance risk: the record can survive while the
reading institution decays. Once that happens, preservation alone
becomes a misleading comfort. Later actors may possess the artifacts of
a system without possessing a stable way to recover what counted as one
action, one authority path, one evidential horizon, or one consequential
outcome. The result is not simply low confidence. It is a legibility
failure
\citep{hutchins1995cognition, scott1998seeing, bowker1999sorting}.

This matters because institutions do not get to wait until understanding
is complete. Courts, operators, regulators, administrators, reviewers,
and affected users all act under epistemic insufficiency. They must
decide what to trust, permit, deny, contest, escalate, or protect before
ontology is settled and often before the relevant object is fully
legible. In distributed AI settings, that pressure is intensified by
authorization asymmetries, delegated action, and runtime heterogeneity.
Recent governance work on authorization propagation, explicit
incompleteness reporting, delegated endorsement, and shared admission
contracts already points in this direction: what matters is not merely
whether an action was possible, but whether later readers can still
recover who the system was acting for, under what standing, with what
evidence, and through what execution path
\citep{tallam2026authorizationpropagation, tallam2026failreport, tallam2026fromcantowould, tallam2026executionenvelopes}.

The paper's central claim is that distributed AI turns institutional
legibility into a first-order governance problem. The issue is not only
that cognition is distributed across components. It is that the ability
to interpret that distribution depends on surrounding institutional
state: who the system was acting for, what authority and boundary
governed the run, what evidence was actually in scope, how the episode
changed over time, and what counted as its consequential resolution.
When those conditions are not preserved, governance fails before more
familiar metaphysical or legal questions have even been reached.

The paper makes five contributions:

\begin{enumerate}
\def\labelenumi{\arabic{enumi}.}
\tightlist
\item
  It introduces the khipu problem for distributed AI: a failure mode in
  which records survive while the institutional capacity to read them as
  one coherent action degrades.
\item
  It distinguishes \textbf{trace retention} from \textbf{interpretive
  continuity}, arguing that artifact preservation is necessary but
  insufficient for governance.
\item
  It argues that many consequential AI outcomes are better understood as
  \textbf{distributed cognitive episodes} than as outputs of bounded
  models.
\item
  It provides a typed account of \textbf{epistemic insufficiency},
  distinguishing missing evidence, ambiguous evidence, and structurally
  unreadable evidence.
\item
  It proposes \textbf{governance workspaces and receipt-bearing
  governance surfaces} as candidate interpretive infrastructure for
  preserving distributed episodes in a governable form.
\end{enumerate}

The aim is therefore not to argue that current systems are already
conscious, nor to offer a general legal theory of personhood or
responsibility. The aim is narrower and, for that reason, more urgent.
It is to show that once cognition is distributed, preserving the reading
institution becomes part of the technical and organizational substrate
of governance itself.

\section{Related Work}\label{related-work}

This paper sits at the intersection of six literatures.

The first concerns distributed and extended cognition. Hollan, Hutchins,
and Kirsh argue that human-computer interaction is often best understood
at the level of distributed systems rather than isolated individuals or
devices \citep{hollan2000distributed}. Hutchins' broader account of
cognition in the wild and Clark and Chalmers' extended-mind argument
both help show that cognitive organization can span artifacts,
environments, and roles rather than terminating neatly at one bounded
agent \citep{hutchins1995cognition, clark1998extended}. Suchman adds a
complementary warning: situated action routinely exceeds the neat
procedural frames imposed on it after the fact \citep{suchman1987plans}.
The present paper borrows these moves, but redirects them toward a post
hoc governance question: what must survive for later institutions to
reconstruct a distributed AI action as one coherent episode?

The second literature concerns legibility, classification, and boundary
objects. Scott treats legibility as an institutional project that
simplifies complex social realities into governable representations
\citep{scott1998seeing}. Star and Griesemer's boundary objects and
Bowker and Star's work on classification show that institutional order
depends on stable representational forms, shared categories, and the
often invisible labor of maintaining them
\citep{star1989boundary, bowker1999sorting}. The khipu problem extends
that line of thought into AI governance. It asks what happens when the
representational artifacts survive but the reading conditions that make
them institutionally actionable decay.

The third literature concerns documentation, provenance, and audit
artifacts. Datasheets for datasets, data statements, model cards,
provenance standards such as PROV-DM, and internal algorithmic auditing
frameworks all argue that trustworthy machine-learning systems require
richer records than model weights or final outputs alone
\citep{gebru2018datasheets, bender2018data, mitchell2019model, moreau2013prov, raji2020closing}.
This paper is sympathetic to that program, but makes a narrower claim.
Richer documentation is necessary yet still insufficient if preserved
artifacts cannot later be read as parts of one bounded consequential
episode.

The fourth literature concerns accountability, contestability, and the
limits of transparency. Kroll et al.~emphasize that accountable
algorithmic systems require more than opaque operational discretion
\citep{kroll2017accountable}. Doshi-Velez and Kim distinguish
interpretability desiderata from institutional adequacy
\citep{doshivelez2017rigorous}. Wachter, Mittelstadt, and Russell show
how post hoc explanation demands arise under legal and administrative
pressure \citep{wachter2017counterfactual}, while Ananny and Crawford
argue that transparency ideals often overstate what seeing a system's
internals actually yields for accountability \citep{ananny2018seeing}.
The present paper adds a different claim: some failures are not merely
failures of explanation, but failures to preserve one stable object of
review at all.

The fifth literature is the technical shift toward tool-using and
multi-agent AI systems. ReAct interleaves reasoning and acting
\citep{yao2022react}, Toolformer trains models to invoke external tools
\citep{schick2023toolformer}, AutoGen treats multi-agent conversation as
a generic application substrate \citep{wu2023autogen}, Generative Agents
and Voyager show how richer long-horizon behavior emerges from
orchestrated memory and tool use
\citep{park2023generative, wang2023voyager}, and MetaGPT foregrounds
explicitly role-structured multi-agent coordination
\citep{hong2024metagpt}. These systems make the paper's governance
question more urgent: once consequential behavior is produced across
models, tools, humans, and policies, model-centric review becomes
increasingly incomplete.

The sixth literature is adjacent governance work on identity,
authorization, runtime admission, and agent mutability. Recent
manuscript-scale work argues that workflow-level authorization
continuity, explicit incompleteness reporting, delegated endorsement,
shared execution-admission contracts, deterministic completeness
benchmarking, and governance under self-modifying identity drift are not
peripheral implementation details, but structural conditions for
governable agentic action
\citep{tallam2026authorizationpropagation, tallam2026failreport, tallam2026partialevidencebench, tallam2026fromcantowould, tallam2026executionenvelopes, tallam2026layeredmutability}.
The present paper does not subsume those contributions. It supplies a
higher-level diagnosis of why such mechanisms matter: they preserve the
institutional reading conditions of distributed cognition.

Relative to these literatures, the paper's contribution is not another
documentation template, another observability layer, or another risk
taxonomy. It is an argument that distributed AI creates a specific
legibility problem for institutions: the possibility that traces survive
while the reading conditions needed for review decay.

\section{The Khipu Problem}\label{the-khipu-problem}

The khipu is useful here not because it provides a decorative analogy
for information loss, but because it marks a specific failure mode that
modern AI governance is structurally vulnerable to. The record can
survive while the reading practice dies. Once that happens, an
institution may possess the artifacts of a system without possessing a
stable way to interpret what those artifacts amount to. The result is
not ignorance in the ordinary sense. It is a more specific condition:
preserved pattern without preserved legibility.

The historical khipu matters because it separates two things that modern
technical systems too often collapse into one. There is, first, the
persistence of encoded structure. There is, second, the persistence of
the social and embodied practice that makes that structure usable. A
surviving khipu can retain information in a material sense while losing
intelligibility in an institutional sense. The cords remain. The knots
remain. The categories may even remain in some partial statistical form.
But the reading institution no longer exists in the way it once did. The
object is preserved. The act of reading is not
\citep{hutchins1995cognition, star1989boundary, bowker1999sorting}.

That distinction is easy to miss in computational settings because
digital systems tempt us to believe that reproducibility of artifact is
equivalent to reproducibility of understanding. We keep the weights,
traces, prompts, tool outputs, authorization metadata, event streams,
and approval records, and assume that enough preserved state will
guarantee retrospective clarity. But the khipu problem suggests a
different possibility. The decisive failure may not be missing data. It
may be the disappearance of the interpretive practice that once made the
data cohere as one meaningful action.

This matters because the relevant interpretive practice is rarely
contained inside the system itself. It lives in operators, institutional
routines, conventions of attribution, assumptions about what counts as
one action, and procedures for relating heterogeneous artifacts back to
one another. In a simple system, those dependencies can be invisible
because they are stable and local. In a distributed system, they become
constitutive. Once cognition is spread across model calls, tools,
retrieval layers, policy gates, human interventions, and runtime state,
the ability to reconstruct what happened depends on more than the
persistence of any single trace. It depends on the persistence of the
institution that knows how to read across them.

The first claim, then, is negative. Auditability is not enough if audit
is understood as artifact retention alone. A complete archive can still
fail to preserve legibility. One can imagine a future distributed AI
system for which every relevant datum has been retained: execution
traces, prompts, model versions, tool invocations, authorization
decisions, uploaded documents, retrieved contexts, and human approvals.
Yet if no one can still answer basic reconstructive questions in a
stable way, the archive has failed in the relevant governance sense.
Which tool action belonged to which cognitive episode? Which human
intervention changed the trajectory of the system, and which merely
observed it? Which retrieved context was constitutive of the final
denial, and which was merely available nearby? Those are not secondary
forensic questions. They are the questions that determine whether later
judgment is even possible
\citep{ananny2018seeing, kroll2017accountable}.

The second claim is constructive. What must be preserved is not merely
output but interpretive continuity. Interpretive continuity means that a
later reader, who may be outside the original operating context, can
still recover what counted as one action, one boundary, one authority,
one episode, or one consequential decision. This is a stronger
requirement than trace retention and a weaker one than full
re-experience. It does not demand perfect replay of the original
cognitive situation. It demands only that the institution retain the
means to read the preserved artifacts as artifacts of a coherent process
rather than as disconnected residues.

\section{From Model to Mesh}\label{from-model-to-mesh}

If the khipu problem is the loss of interpretive continuity, the next
question is what sort of object later institutions are trying to read.
In older governance frames, the answer was usually a bounded model or a
bounded software component. That answer is becoming less adequate.
Consequential outputs increasingly emerge from orchestrated episodes
across models, retrieval systems, tools, policy services, runtime
managers, and humans in the loop
\citep{yao2022react, schick2023toolformer, wu2023autogen, hong2024metagpt}.

This shift matters because the action to be governed is no longer
identical to any one component. A retrieval module may determine
evidential scope without ever issuing the final answer. An approval
service may determine whether an action is admissible without
contributing any linguistic output. A runtime manager may determine
which model family or policy version actually executed the consequential
step. Human intervention may redirect the episode while leaving only a
small local trace. The visible output is real, but it is no longer a
sufficient proxy for the process that produced it.

That is why recent work on authorization propagation, delegated
endorsement, and shared execution-admission contracts matters for more
than local security or backend design. These works all treat identity
continuity, standing, boundary state, and runtime admission as
first-class governance surfaces because modern agentic behavior already
exceeds the simple bounded-model frame
\citep{tallam2026authorizationpropagation, tallam2026fromcantowould, tallam2026executionenvelopes}.
They do not solve the khipu problem by themselves, but they identify
some of the structural relations a later institution must be able to
recover.

The right unit for many governance questions is therefore neither the
model nor the whole institution, but a \textbf{distributed cognitive
episode}: a bounded span of coordinated activity across models, tools,
humans, context, and policy that produces a consequential outcome. Such
an episode may still be temporally short and operationally local, but it
is not reducible to a single component without loss of meaning. What
matters is not that many subsystems were present. What matters is that
the final act depends on the structure of their coordination.

\subsection{Minimal Formalization}\label{minimal-formalization}

Let a distributed cognitive episode be represented as:

\[
e = \langle i, q, a, b, s, \rho, o, T \rangle
\]

where \texttt{i} is episode identity, \texttt{q} is requester identity,
\texttt{a} is the effective authority path, \texttt{b} is the governing
boundary regime, \texttt{s} is the evidential scope partition,
\texttt{\textbackslash{}rho} is runtime lineage, \texttt{o} is
consequential outcome identity, and \texttt{T} is the preserved trace
set for the episode.

For governance purposes, the evidential scope partition \texttt{s} is
not just a bag of retrieved artifacts. It is usefully treated as a
bounded partition among evidence used, evidence available but unused,
and evidence structurally inaccessible under the governing boundary.
That is exactly the distinction later institutions need when deciding
whether a system acted with full access, partial but acceptable access,
or structurally incomplete access
\citep{tallam2026failreport, tallam2026partialevidencebench}.

This notation does not pretend to solve the ontology of agency. It
isolates the minimum relational structure that later institutional
reading must often recover. The point is not that every system already
exposes these fields explicitly. The point is that governance pressure
arises when something functionally like them is needed for review but
has not been preserved in stable form.

\section{Institutional Legibility}\label{institutional-legibility}

If the previous section shifts the governed object from model to
distributed episode, this section shifts the governance problem from
retention to legibility. Once the relevant act is a distributed
cognitive episode, the central question is no longer whether the
institution possesses enough traces in some abstract sense. The question
is whether those traces can still be read in a stable way by the kinds
of actors who will later have to rely on them. Institutional legibility
names that condition.

Legibility here should be understood in a strict institutional sense. It
is not synonymous with mere interpretability, observability, or
explainability. A system can be locally explainable while remaining
institutionally illegible. It can surface a persuasive rationale while
obscuring which authority path, scope boundary, or approval condition
made the action admissible. It can preserve extensive logs while forcing
later review to depend on operator memory or tacit architectural
knowledge. In those cases, the traces are present but the reading
conditions are weak \citep{scott1998seeing, ananny2018seeing}.

We can express the difference between trace retention and interpretive
continuity with a simple reading function. Let
\texttt{\textbackslash{}mathcal\{R\}\_I(T(e))} denote the attempt by
institutional reader \texttt{I} to recover a governable account of
episode \texttt{e} from its preserved traces. Then define interpretive
continuity as:

\[
IC_I(e) = 1
\]

iff the institutional reader can recover episode identity, requester
truth, authority path, boundary regime, evidential scope, runtime
lineage, and consequential outcome from preserved artifacts without
relying on tacit local memory. The point is not that every reader must
recover everything equally well. The point is that the institution must
preserve a stable procedure by which those relations remain
reconstructable.

\begin{figure}[!htbp]
\centering
\includegraphics[width=0.76\linewidth]{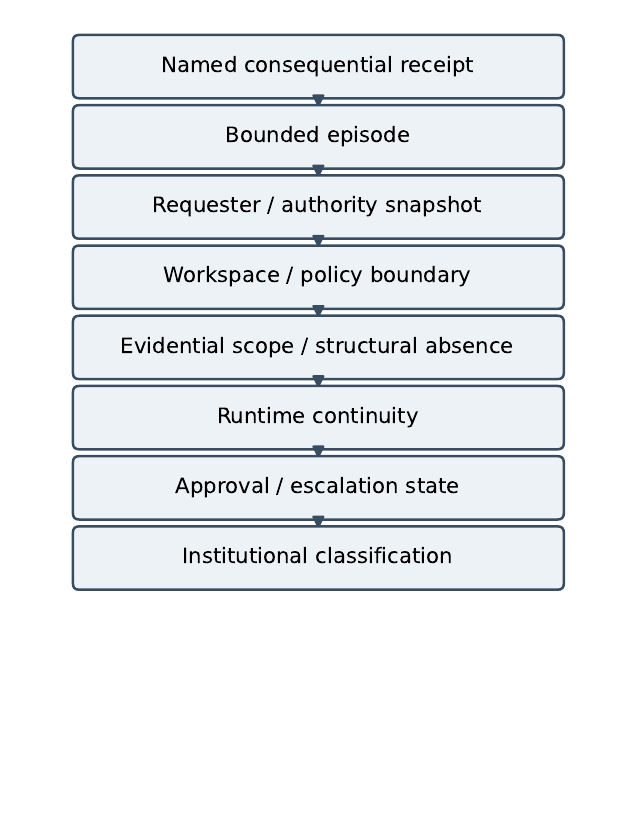}
\caption{A bounded interpretive review path for a consequential distributed episode. The point is not the specific receipt format, but the existence of a stable path from institutional outcome back through episode identity, authority, boundary, evidential scope, runtime continuity, and approval state.}
\label{fig:review-path}
\end{figure}

This requirement is practical rather than metaphysical. Consider a
consequential denial under partial visibility. A reviewer later asks
whether the denial was appropriate. The answer cannot come from the
output text alone. The reviewer may need to know whether a relevant
annex or customer record was in scope, whether the action occurred under
constrained standing, whether a human approval was required, whether the
runtime changed between draft and final action, and whether the correct
classification is ordinary error, acceptable partiality, or unreadable
action. Figure\textasciitilde{}\ref{fig:review-path} shows the minimum
review path needed to answer that kind of question without turning
oversight into archaeological reconstruction.

The review path in Figure\textasciitilde{}\ref{fig:review-path} is meant
to emphasize that interpretive continuity is not just richer telemetry.
It is a preserved sequence of institutional reading steps. A named
consequential receipt anchors the case. A bounded episode keeps
heterogeneous traces from collapsing into ambient system activity.
Requester and authority snapshots preserve standing. Boundary state
preserves what the system could and could not legitimately see. Runtime
continuity preserves what actually executed the consequential step.
Approval and escalation state preserve what institutional gating should
have happened. If these relations are not preserved, then later readers
may still have traces, but they no longer have a stable case file.

This is the place where the paper can stop speaking only negatively. If
institutional legibility is the condition under which a distributed
cognitive episode remains governable, then the design task is to build
environments that preserve reading conditions as first-class
infrastructure. Such environments do not merely host interaction. They
preserve requester truth, boundary truth, role history, runtime
continuity, approval history, evidential scope, and outcome identity in
a way that later readers can still use.

\section{Epistemic Insufficiency and
Action}\label{epistemic-insufficiency-and-action}

Institutions rarely enjoy the luxury of waiting until reading conditions
are perfect. They act under epistemic insufficiency. The term is meant
to be narrower than ignorance and broader than uncertainty. An
institution faces epistemic insufficiency when it lacks enough settled
understanding to classify a consequential episode cleanly, yet cannot
defer action without cost. The insufficiency may concern the relevant
facts, the meaning of the facts, the unit to which they should be
attributed, or the ontology of the object being governed. In distributed
AI systems, all four can fail at once
\citep{suresh2019framework, wachter2017counterfactual}.

What matters first is to distinguish different forms of insufficiency,
because they imply different action norms. The simplest case is
\textbf{missing evidence}. Here the relevant categories are stable and
the missing information is, in principle, the sort of thing the
institution knows how to use. A required document was never retrieved,
an approval record was not preserved, a runtime handoff was not logged,
or a user-side event was not captured. The institution may still know
exactly what kind of fact it lacks.

A harder case is \textbf{ambiguous evidence}. Here the institution has
preserved artifacts, but their meaning is contested or underdetermined.
A tool log exists, but it is unclear whether it was decisive or
peripheral. A retrieved document was present, but it is unclear whether
it was constitutive of the final answer. A human intervention occurred,
but it is unclear whether it changed the course of action or merely
observed it. Ambiguous evidence is not absence. It is the failure of
clear mapping from evidence to institutional meaning.

The hardest case, and the one most central to this paper, is
\textbf{structurally unreadable evidence}. In this condition, the issue
is not merely that some fact is missing or that some preserved artifact
is open to competing interpretations. The issue is that the system has
not preserved the reading conditions required to make the artifacts
cohere as one governable episode at all. The institution lacks not just
a fact, but a stable way to say what kind of fact would have settled the
matter. It may have a mass of traces without a defensible procedure for
telling what counted as one action, one requester, one authority path,
one evidential scope, or one outcome identity
\citep{tallam2026failreport, tallam2026partialevidencebench}.

\begin{table}[!htbp]
\centering
\small
\caption{Typed epistemic insufficiency for distributed AI governance.}
\label{tab:insufficiency}
\begin{tabular}{|p{1.95cm}|p{2.1cm}|p{2.55cm}|p{2.75cm}|p{2.75cm}|}
\hline
\textbf{Insufficiency type} & \textbf{What is preserved} & \textbf{What is missing or failing} & \textbf{Institutional symptom} & \textbf{Appropriate action norm} \\
\hline
Missing evidence & Stable episode categories and partial records & A required fact or artifact is absent & The reviewer knows what kind of record needs recovery & Recover, supplement, or narrow confidence \\
\hline
Ambiguous evidence & Episode-linked artifacts are present & The meaning or causal role of artifacts is contested & The reviewer has a case but not a settled interpretation & Review, contest, escalate, or preserve disagreement \\
\hline
Structurally unreadable evidence & Raw traces and local artifacts may be present & The reading conditions that bind them into one governable episode are missing & The reviewer lacks a stable object of institutional judgment & Block, narrow, refuse, or route to richer review before treating the outcome as settled \\
\hline
\end{tabular}
\end{table}

These distinctions matter because institutions should not respond to
them in the same way. Missing evidence often calls for recovery,
supplementation, or narrower confidence. Ambiguous evidence often calls
for interpretation, review, contestability, or explicit disagreement
procedures. Structurally unreadable evidence calls for something
stronger: blocking, escalation, narrowed permissions, provisional
treatment, or explicit refusal to pretend that the action is fully
classifiable. If these cases are collapsed into one generic bucket of
uncertainty, institutions tend either to overreact to ordinary ambiguity
or, worse, underreact to unreadability by treating it as merely another
confidence score.

We can write that action pressure compactly as:

\[
Unreadable_I(e) \iff \neg StableRecover_I(e)
\]

where \texttt{StableRecover\_I(e)} means that institution \texttt{I} can
reconstruct the governance-relevant tuple for episode \texttt{e} as one
coherent case. The operational action norm can then be sketched as:

\[
Norm(e) =
\begin{cases}
\text{recover/supplement} & \text{missing evidence} \\
\text{review/contest/escalate} & \text{ambiguous evidence} \\
\text{block/narrow/refuse} & \text{structurally unreadable evidence}
\end{cases}
\]

The point of the notation is not to formalize institutional judgment
away. It is to prevent a category error. Confidence language alone
cannot represent the difference between ordinary uncertainty and
structural unreadability. That is why fail-and-report style controls and
deterministic completeness benchmarks matter. They force systems to
treat structurally constrained visibility as a first-class governance
event rather than as an invisible background condition
\citep{tallam2026failreport, tallam2026partialevidencebench}.

\section{Governance Workspaces as Interpretive
Institutions}\label{governance-workspaces-as-interpretive-institutions}

If institutions need a stable way to read distributed cognitive
episodes, then they need more than model access and more than generic
audit logs. They need an environment that preserves the episode's
institutional shape for later readers. A \textbf{governance workspace},
understood strongly, is one candidate institutional form.

The phrase does not denote a particular product surface. It denotes a
bounded institutional container in which distributed action is
accumulated, named, scoped, reviewed, and resolved. Such a workspace
preserves more than conversation. It preserves who the system was acting
for, what authority path was in force, what evidence was in or out of
scope, what runtime family executed the consequential step, what
approval or escalation state applied, and what counted as the final
institutional outcome. In other words, it preserves the reading
conditions that later make the episode governable.

This is where adjacent governance work becomes concretely relevant.
Authorization propagation treats standing and boundary continuity as
workflow-level requirements rather than per-call afterthoughts
\citep{tallam2026authorizationpropagation}. From Can to Would separates
what an agent \emph{can} do from what it is normatively acting
\emph{for} on behalf of a requester \citep{tallam2026fromcantowould}.
Execution envelopes identify admission-time objects that preserve who
requested what kind of execution and what the backend ultimately granted
\citep{tallam2026executionenvelopes}. Taken together, these are not
merely local control features. They are ingredients of an interpretive
institution.

The value of a governance workspace is therefore not that it stores more
data. It is that it preserves \emph{bounded institutional objects}. A
consequential denial should not survive only as a pattern implicit in
logs. It should survive as a receipt-bearing event linked to episode
identity, requester truth, authority path, boundary state, evidential
scope, and review status. A blocked action should not survive only as a
transient exception. It should survive as a named institutional event. A
human escalation should not survive only as a timestamped message. It
should survive as part of the governed episode's resolution path.

This reframes a common enterprise question. The issue is not merely
whether a system offers collaboration, approvals, or provenance. The
issue is whether those surfaces preserve a later-readable case file. If
they do not, then the institution may still have conversation history
and telemetry without having a governable action object. If they do,
then later review becomes something stronger than log search: it becomes
structured institutional reading.

\section{Coordination Meshes and the Unit of
Governance}\label{coordination-meshes-and-the-unit-of-governance}

The governance-workspace argument ends with a harder question than it
begins with. If a workspace preserves a distributed episode in
institutionally readable form, what exactly is the thing being governed?
Is it the model that generated the visible response? The runtime that
executed the episode? The bounded run itself? The governance workspace
that preserved the episode's identity? The larger coordination mesh of
models, tools, policies, context systems, and humans through which the
episode emerged? Or, in some cases, the broader human-machine
arrangement that delegated and constrained the result?

There is no serious way to answer that question by insisting in advance
that only one unit can ever matter. Different governance judgments care
about different structures. A model-level safety question may genuinely
target one model family. A deployment review may target the runtime. A
contestation or appeal may target the bounded episode. A policy audit
may target the workspace and its boundary regime. A systemic governance
review may target the coordination mesh or the broader delegation
arrangement. The right lesson is not pluralism without discipline. It is
\textbf{situational governance}: choose the smallest unit that preserves
the causal and institutional structure relevant to the judgment at hand.

\begin{figure}[!htbp]
\centering
\includegraphics[width=0.90\linewidth]{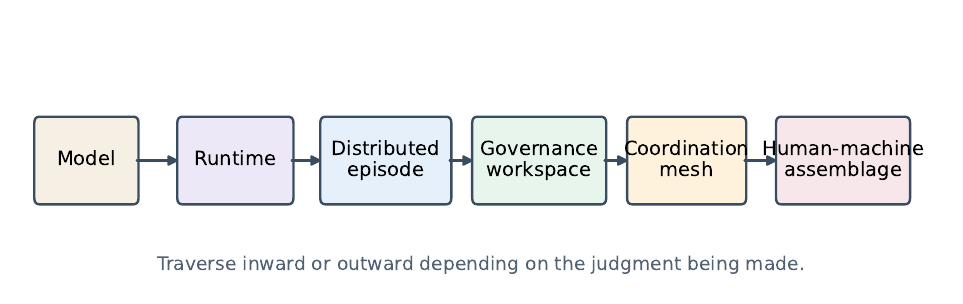}
\caption{Question-relative units of governance for distributed AI. Later review may need to traverse among these units rather than treating any single layer as universally authoritative.}
\label{fig:units}
\end{figure}

Figure\textasciitilde{}\ref{fig:units} depicts this as a traversable
stack rather than a single privileged layer.

The need for traversability becomes especially clear once the governed
object itself drifts over time. Work on layered mutability shows that
persistent agents can accumulate identity-relevant change across memory,
self-narrative, adaptation, and downstream coupling even when no one
visible update looks catastrophic \citep{tallam2026layeredmutability}.
Under those conditions, later governance cannot assume that the same
unit remains equally appropriate across time. Some questions will target
local behavior. Others will target the mutation path that made later
behavior possible. Unit choice is therefore not only question-relative.
It can also be temporally unstable.

This situational view clarifies a common confusion in debates about AI
agency. People often argue past each other because they are assigning
agency-talk to different units without saying so. One person is speaking
about model-level generativity. Another is speaking about episode-level
consequential action. Another is speaking about workspace-level
delegated workflow. Another is speaking about mesh-level coordination.
Another is speaking about the broader socio-technical arrangement. The
result is not just conceptual disagreement. It is unit slippage. Once
the unit of governance is made explicit, many of these disputes become
at least better posed.

The practical upshot is simple. Governance should treat the unit of
analysis as question-relative but reconstruction-dependent. The correct
unit is the one that preserves enough of the distributed episode to
support the judgment being made, and the system must preserve enough
connective tissue that later readers can justify that choice. Cross-unit
traversability is therefore not an optional observability luxury. It is
part of what makes situational governance intellectually honest.

\section{Design Implications}\label{design-implications}

The design target for distributed AI systems has been mis-specified. The
usual target is some mixture of capability, safety, observability, and
compliance. Those remain necessary, but they are not enough. A system
can be capable, apparently safe, richly instrumented, and formally
auditable while still failing the central governance requirement
developed in this paper: preserving a distributed cognitive episode in a
form that later institutions can still read. The design problem is
therefore not just to log more and explain more. It is to preserve
interpretive continuity.

That shift produces a different set of requirements.

The first is to preserve \textbf{episode identity} rather than only
component activity. A system should not merely record that a model was
called, a tool was invoked, a document was retrieved, a policy gate
fired, and a human approval occurred. It should preserve which of those
events belonged to the same consequential episode and under what shared
institutional terms they were linked. Without stable episode identity,
later readers are forced to infer composition from raw adjacency.

The second requirement is to preserve \textbf{requester truth and
authority truth} as first-class state. Many systems record requests
while treating the deeper question of standing as ambient. That is a
mistake. A later institution often needs to know not just what was
asked, but who the system was acting for, what role or delegation path
authorized the request, and whether that authority shifted during the
episode. Requester and authority truth should therefore not live only in
operator memory or weakly linked metadata
\citep{tallam2026authorizationpropagation, tallam2026fromcantowould}.

The third requirement is to preserve \textbf{boundary truth}.
Distributed cognition unfolds inside scoping structures: tenants,
projects, policy domains, credential surfaces, data compartments,
execution environments, and governance workspaces. Those boundaries are
not merely administrative. They define evidential access, tool
availability, approval requirements, and the range of possible plans.
Systems should therefore preserve which boundaries were in force at the
moment of consequential action and how those boundaries changed, if they
changed at all.

The fourth requirement is to preserve \textbf{evidential scope,
including structural absence}. It is not enough to know what the system
retrieved or cited. For many governance questions, one must also know
what was available but unused and what was unavailable because of
authorization or boundary constraints. A design that records only what
was used but not what could or could not have been seen makes
incomplete-action cases harder to classify later
\citep{tallam2026failreport, tallam2026partialevidencebench}.

The fifth requirement is to preserve \textbf{consequential receipts}
rather than relying on unbounded trace review. A mature system should
emit stable institutional events for admissions, denials, approvals,
escalations, blocked actions, partial completions, and terminal
outcomes. These receipts should carry the institutional terms that make
them later readable: requester, role, scope, runtime, evidence horizon,
and named status taxonomy. The point is not bureaucratic ornament. It is
to give later institutions bounded objects of judgment
\citep{tallam2026executionenvelopes}.

The sixth requirement is to preserve \textbf{degraded-path legibility}.
Systems are often designed to be most readable on their nominal success
path and least readable precisely where governance pressure is highest.
Failed actions, denied requests, ambiguous approvals, missing evidence,
runtime handoffs, and human overrides are left scattered across logs
because they are operational exceptions. From the perspective of
institutional legibility, that is backwards. These paths are exactly
where later readers need the strongest reconstruction support.

The seventh requirement is to preserve \textbf{cross-unit
traversability}. A reviewer should be able to start from a model output
and recover the runtime; from the runtime, recover the episode; from the
episode, recover the workspace; from the workspace, recover the standing
coordination mesh conditions; and from there understand what broader
human delegation pattern mattered. A system that optimizes for one
privileged cut alone will make at least some classes of later governance
artificially hard \citep{tallam2026layeredmutability}.

The eighth requirement is to treat \textbf{provenance as active
infrastructure} rather than passive aftercare. Provenance is often
introduced late, as if it were a reporting layer that can be attached
once the ``real'' system exists. The khipu problem suggests the
opposite. If later interpretation depends on stable links among
distributed artifacts, then provenance has to be part of the
architecture that constitutes the episode in the first place
\citep{moreau2013prov, raji2020closing}.

The ninth requirement is to \textbf{fail closed on unreadable
consequential action}. If a system cannot preserve enough of an episode
to support later institutional reading, then certain classes of
consequential action should not proceed as though nothing is wrong. This
does not mean every incomplete or partially legible action must halt. It
means the system should distinguish between ordinary uncertainty and
structural unreadability. Where later accountability, appeal, or
reconstruction will matter, unreadable action should trigger blocking,
escalation, narrower permissions, or explicit partial-status handling
rather than silent completion \citep{tallam2026failreport}.

The tenth requirement is to design for \textbf{future outsiders}, not
only present operators. Much current observability and audit design
assumes a cooperative reader who already knows the architecture and
shares the local vocabulary. That assumption is too weak. The relevant
future reader may be a new team, a customer, a regulator, a court, or an
affected party. Systems should therefore preserve enough explicit
structure that such readers can recover what happened without depending
on tacit institutional memory.

Taken together, these requirements define a different design posture.
Instead of treating distributed cognition as a performance problem
wrapped in optional governance surfaces, one treats governability as
part of the core systems contract. A good system is not merely one that
acts effectively. It is one that preserves the terms under which its
actions can later be reconstructed, contested, attributed, constrained,
and, when necessary, refused.

\section{Conclusion}\label{conclusion}

The central claim of this paper is simple. As AI systems become more
distributed, the hardest governance problem is not only what they can
do. It is whether institutions can still read what they have done. The
relevant failure mode is not exhausted by opacity, low confidence, or
missing logs. It is the khipu problem: the possibility that the record
survives while the reading practice decays, leaving later institutions
with preserved artifacts but no stable way to recover one governable
object from them.

That claim required several shifts. The first was from bounded model to
distributed episode. Consequential behavior increasingly emerges not
from one artifact but from coordinated activity across models, tools,
retrieval layers, runtime policy, authorization boundaries, and human
intervention. The second was from retention to legibility. It is not
enough to preserve traces if those traces cannot still be read by later
operators, regulators, reviewers, courts, or affected users. The third
was from generic uncertainty to typed epistemic insufficiency. Missing
evidence, ambiguous evidence, and structurally unreadable evidence are
different conditions and should trigger different action norms. The
fourth was from abstract diagnosis to concrete institutional form. If
distributed episodes are to remain governable, they need environments
that preserve requester truth, boundary truth, authority continuity,
evidential scope, runtime lineage, and bounded receipts for
consequential action.

The argument also resisted the temptation to force one final answer to
the question of what is being governed. Sometimes the right unit is the
model. Sometimes it is the runtime, the episode, the governance
workspace, the coordination mesh, or the broader human-machine
arrangement. What matters is that systems preserve enough connective
tissue for later institutions to justify that choice without fiction.
Situational governance is not a retreat from rigor. It is what rigor
looks like when the governed object is structurally distributed.

This is why the khipu problem is a governance problem before it is a
metaphysical one. One need not settle the strongest questions about AI
agency, consciousness, or moral standing to face the practical
challenge. Institutions already have to classify, trust, constrain,
review, and sometimes protect systems whose relevant behavior exceeds
clean representational frames. If those systems cannot preserve the
terms under which their actions remain readable, then governance fails
before those further philosophical questions can be handled responsibly.
The coming problem is not only distributed intelligence. It is
distributed intelligence that must remain institutionally legible under
real conditions of action, contest, and time.

\bibliography{references.bib}

\end{document}